\documentclass[aps,prl,twocolumn,superscriptaddress,floatfix]{revtex4-1}
\usepackage{graphicx}
\usepackage{amssymb}
\usepackage{epsfig}
\usepackage{amsmath}
\usepackage{datetime}
\usepackage[dvipsnames]{xcolor}
\usepackage{tabularx}
\usepackage[normalem]{ulem}
\usepackage{hyperref}
\hypersetup{
	colorlinks=true,       
	linkcolor=blue,          
	citecolor=blue,        
	filecolor=magenta,      
	urlcolor=blue           
}
\newcommand{\red}{\color{red}}

\newcommand{\etal}{\textit{et al.}}

\newcolumntype{b}{X}
\newcolumntype{s}{>{\hsize=.5\hsize}X}
\begin{document}

\title{
Activity-driven sorting, approach to criticality \\and turbulent flows in dense persistent active fluids }

\author{Suman Dutta}
\email{Corresponding Author, Email: suman.dutta@icts.res.in}
\affiliation{Amrita School of Artificial Intelligence, Coimbatore, Amrita Vishwa Vidyapeetham, India}
\affiliation{National Centre for Biological Sciences-TIFR, Bengaluru 560065, India}
\affiliation{International Centre for Theoretical Sciences-TIFR,
Bengaluru 560089, India}
\author{Pinaki Chaudhuri}
\affiliation{The Institute of Mathematical Sciences, Chennai, India}
\author{Madan Rao}
\affiliation{Simons Centre for the Study of Living Machines, National Centre for Biological Sciences-TIFR, Bengaluru 560065, India}
\author{Chandan Dasgupta}
\affiliation{International Centre for Theoretical Sciences-TIFR,
Bengaluru 560089, India}
\affiliation{Indian Institute of Sciences, Bengaluru, India}

\date{~\today}

 \begin{abstract}

We show that dense active fluids comprising interacting particles with \textcolor{black}{persistent self-propulsion are driven to a non-equilibrium steady state consisting of co-moving particles with co-aligned active forces. This velocity and force sorting appears to be associated with a critical state where the length scales associated with spatial correlations of the velocity and the propulsive force grow with system size.} 
At large system sizes, these growing velocity domains are accompanied by the appearance of dynamic macroscopic voids in the steady state, associated with large density fluctuations.
The \textcolor{black}{dynamics} of the macroscopic voids drives a new kind of turbulent state.
 
 
\end{abstract}                   

\maketitle



In this paper, we report on the unusual liquid state properties of persistently driven active athermal fluids in the dense state. 
It is recognized that many of the unconventional features of active matter are consequences of microscopic time reversal symmetry (TRS) breaking \cite{trs1,trs2, active-review1, active-review2}. There have been several recent studies on the collective dynamics of dense active particles as a function of activity and persistence time \cite{dam1,ni2013,dam3,berthier2014,berthier2017,mandal2016,mandal2017,dam5,dam7}. Dense active matter at large persistence time and intermediate forcing~\cite{dam8} shows intermittent stress build-up and sudden relaxation through plastic events leading to turbulence ~\cite{dam8,keta2022,keta2023,keta2024}.  In the limit of infinite persistence time, the particles show a jamming transition~\cite{dam8,liao2018,yang2022} with forces channeled along chains or rays~\cite{dam9}. Here, we study the unjammed state of the persistent active fluid at higher forcing, using large-scale simulations.




\begin{figure*}[htb]
\includegraphics[width=1.9\columnwidth]
{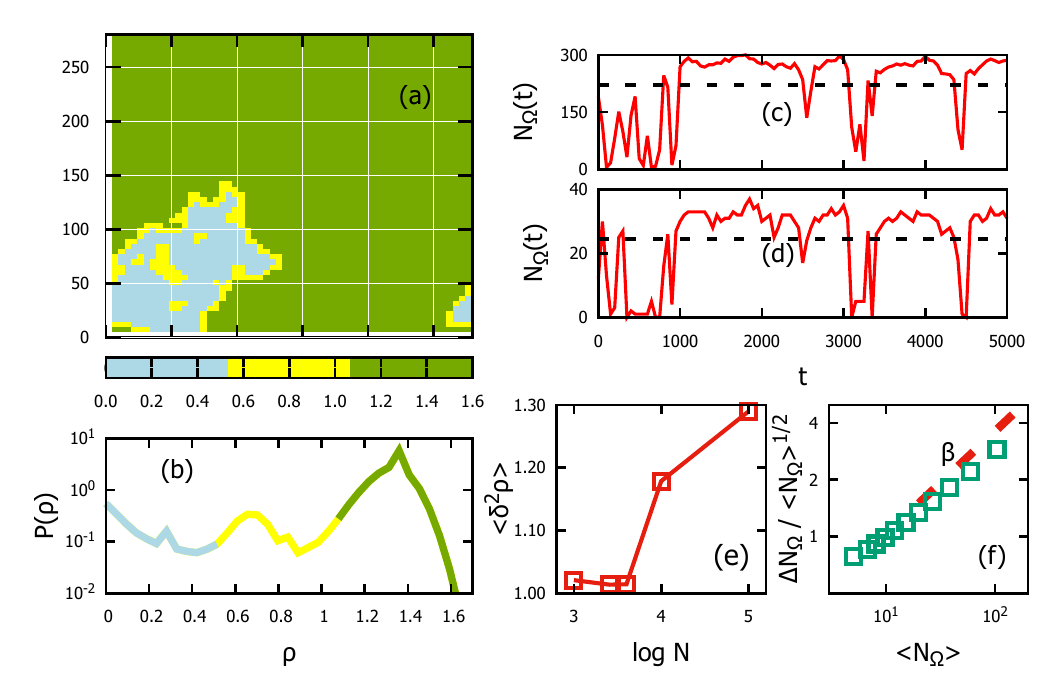}
\caption{(a)\,Spatial map of particle density $\rho(x,y)$ at the steady state, using a coarse-graining scale $\Omega=4.73$  for $N=10^5$ particles. This clearly shows the appearance of macroscopic voids. (b)\,The corresponding steady state probability distribution of the local density $\rho$ shows multi-modality, with the three peaks associated with the void, the interface and the bulk. (c-d)\,Typical time series of $N_{\Omega}$, the number of particles in box of sizes $\Omega(=13.74(c)$ and $4.73(d))$ taken from a large system with $N=10^5$. The respective mean of the timeseries is marked by the dashed line. The large fluctuations and sudden drops and rises reflect the transit of large voids across the box. (e)\,Variance of the local density $\delta^2 \rho$, calculated from the moments of $P(\rho)$ and its dependence on $N$. As elsewhere, the data are extracted from $50-100$ independent runs with independent initial conditions for each $N$. (f)\,Giant number fluctuations displayed as a log-log plot of the standard deviation $\triangle N_{\Omega}$ versus the mean $\langle N_{\Omega}\rangle$ for different box sizes $\Omega$ taken from a large system $N=10^5$. The fit at large $\langle N_{\Omega}\rangle$, shows $\Delta N_{\Omega} \sim   \langle N_{\Omega}\rangle^{\frac{1}{2}+\beta}$, with $\beta\simeq 0.5$. The dashed line shows the best fit to the large-$N_\Omega$ data with $\beta = 0.456$ (see SI for related discussion).}
\label{dens_fluct}
\end{figure*}

Our main result is that persistent activity generically drives the liquid to a non-equilibrium critical state with correlations that grow with system size $L$. Starting from initial conditions where the particle velocities and propulsion forces are randomly assigned whilst maintaining isotropy, the system evolves into domains of co-moving particles with co-aligned propulsive forces, separated by regions of high vorticity. For small system sizes, the size of these velocity domains scales with system size. This velocity and force sorting 
is a consequence of TRS-breaking arising from collisions and self-propulsion and implies that in contrast to equilibrium liquids, the joint distribution of positions and velocities do not decouple. 
At larger system sizes, the system also exhibits strong density inhomogeneity with macroscopic voids and anomalous density fluctuations. This is accompanied by
giant number fluctuations where the standard deviation of the number of particles $N_\Omega$ in a sub-region of size $\Omega$ scales as $\Delta N_\Omega \propto {N_\Omega}^{\frac{1}{2}+\beta}$, where $\beta \simeq 0.5$. The motion of the macroscopic voids appear to drive a kind of large-scale turbulence, characterized by strong dynamical heterogeneity.

Our results follow from extensive computer simulations of a 2-dimensional athermal system of binary Lennard-Jones particles \cite{koboriginal} of mass $m$, each of which is subjected to a propulsion force ${\bf f}_{i}=f{\hat n}_{i}\equiv f (\sin{\theta_i}, \cos{\theta_i})$ \cite{dam8,rituporno2}, using LAMMPS, \cite{lammps} \textcolor{black}{following an equation of motion for an active particle at position $\bf r_{i}$, within a medium of damping $\Gamma$}: $m{\ddot{\bf r}}_{i}=-\Gamma{\dot{\bf r}}_{i} +\sum^{N}_{j \ne i=1}{\bf f}_{ij}+{\bf f}_{i}\,$ where, initially, the active forces are assigned randomly maintaining isotropy, $\langle{\bf f}_{i}\rangle =N^{-1} \sum_{i=1}^{N}{\bf f}_{i}= 0$, \textcolor{black}{at the complete absence of thermal fluctuation}. In our persistently active system, the initially assigned propulsive forces are not allowed to change and isotropy is maintained at all times. This is equivalent to having particles endowed with an exponentially correlated noise with 
infinite persistence time~\cite{active-saikat}. Details of the model parameters and simulations can be found in the Supplementary Information (SI) Note 1.




Our athermal simulations are carried out in a regime where the propulsive forces are large enough, $f \gg f_{c}\approx 1.65$ (for $N=10^3$)~\cite{dam8},  to ensure that the system is an unjammed fluid. Close to and below this threshold force, the system shows intermittent jamming and stress release via plastic events that are highly sensitive to system size. Here we focus on liquid state properties. We thus choose 
$f=3$ for particle numbers ranging between $N=10^2-10^6$. In a later publication we will address the physics of jamming as one \textcolor{black}{quenches} the propulsion force $f$~\cite{suman0}.

For small system sizes $N\leq10^3$, a static snapshot of the structure of the persistently driven liquid at the steady state is indistinguishable from the passive dense liquid (SI Fig.\eqref{config}). This is reflected in  the radial distribution function $g(r)$ (SI Fig.\eqref{rdf}). The probability distribution of the density, $P(\rho)$ however, show a multi-modal distribution (SI Fig.\eqref{rdf}).

\textcolor{black}{On increasing $N (\geq 10^4)$, the system surprisingly show the appearance of large voids (Fig.1(a)), suggesting cavitation or a gas-liquid coexistence, in the form of motility-induced phase separation, even in persistent states} ~\cite{cates2015} in which the density in the ``gas'' phase is close to zero. This is reflected in the steady state density distribution  $P(\rho)$ which shows non-Gaussian features and multi-modality (Fig.1(b)) with the additional peak at $\rho=0$ representing the void regions. SI Movie 1 shows that the large density inhomogeneities are dynamic and grow with system size (SI Fig.\eqref{number_density}). The steady state density variance \textcolor{black}{grows with the appearances of larger cavities, beyond a limit} (Fig.1(d)). 
Concomitantly, 
the system exhibits anomalously large number fluctuations, as seen from the time series of the number of particles computed within boxes of different sizes $\Omega$ (Fig.1(c)). \textcolor{black}{We also note} the sudden drops and rises as the void sweeps through the sampling box $\Omega$. The standard deviation of the number of particles for each $\Omega$, is related to the mean number by $\Delta N_{\Omega} \sim \langle N_{\Omega}\rangle^{\frac{1}{2}+\beta}$, where $\beta \simeq 0.5$ (Fig.1(e)). 
This is a simple consequence of the large variations of $N_\Omega(t)$ (Fig.1(c)) which fluctuates between a large value corresponding to the local density close to the third peak in Fig.\,1(b) and a value close to zero. 
We note that a 
computation of $\Delta N_{\Omega}$ versus $\langle N_{\Omega}\rangle$ for fixed $\Omega$ but varying $N$ shows a different trend owing to the absence of voids at smaller $N$; nevertheless the asymptotic behavior at large $N$ is similar to Fig.1(e) (see SI Fig.\,\eqref{giant_number}).
The giant number fluctuations found here are \textcolor{black}{likely to have a different microscopic origin than that described in} Ref.~\cite{adititonersriram} which are generated by large particle currents induced by Goldstone fluctuations of the polarization field in a nematically ordered active fluid.




This persistently driven active fluid exhibits an interesting self-organization of the particle velocities and propulsion forces.
Analysis of the spatial maps 
of the orientations of the coarse-grained velocity and propulsive force,
   $\Theta = \tan^{-1}v_y/v_x$ and  $\Phi = \tan^{-1}f_y/f_x$, respectively (Fig.2 (a) and (b)), shows that the system reorganizes into domains of co-moving particles with co-aligned  propulsion forces. Figs. 2 (a-c) reveal that the velocity and propulsion force maps show appreciable correlations. As shown in Fig. 2(c), the orientation $\theta$ of the time-averaged velocity of a particle in the steady state is close to the orientation $\phi$ of the propulsion force acting on it, indicating that each particle acquires a non-zero average velocity in the direction of the constant propulsion force. Note that the propulsion forces show a weaker co-alignment than the velocities in a given domain. We find that the velocity domains are separated by regions of high vorticity (Fig. 2(d)).


  
  This sorting of the velocity and the propulsion force is a consequence of clustering and co-alignment following low-angle collisions of particles whose propulsion forces are nearly co-aligned.
 Velocity sorting implies strong correlations between position and velocity -- the joint distribution of the two does not decouple.

The size $\xi$ of the domains of co-moving and co-aligned particles 
 can be obtained from the equal-time velocity-velocity $C_{vv}$ and force-force $C_{ff}$ correlation functions in the steady state (Fig. 3(a), also see SI Note 4). These length scales, obtained as the values of $r$ at which the correlation functions cross zero, grow as $\sqrt{N}$ for small $N$ indicating growing critical correlation lengths (Fig. 3(b)) that would diverge in the thermodynamic limit. The data for the largest system ($N=10^5$) fall below the $\xi \propto \sqrt{N}$ line obtained from a fit to the data for smaller $N$. This suggests that the presence of prominent voids in large systems may lead to an eventual saturation of $\xi$ as $N \to \infty$. However, we cannot confirm this possibility from the present data. The occurrence of strong velocity correlations in dense active systems with large persistence time has been observed in several studies~\cite{henkes2020,szamel2021,caprini2020a,caprini2020b,kuroda2023}. In these studies, the length scale of velocity correlations is found to increase with increasing persistence time, suggesting a divergence in the limit of infinite persistence time considered here. However, the presence of voids found in our simulations for large $N$ was not considered in these studies. To our knowledge, the presence of long-range spatial correlations of the propulsion forces has not been reported earlier. These correlations require a higher degree of self-organization in which particles with similar directions of propulsion forces, which may be separated by large distances in the initial random state, come close to one another to form large clusters in which all the particles have similar directions of propulsion forces. The alignment of velocities can occur via local collisions without involving motion of particles over large distances.
The mean kinetic energy per particle $<E_K>$ saturates to a finite value (SI Fig.\,\eqref{ke}), with a distribution $P(E_K)$ that 
is non-Gaussian with a positive skewness(SI Fig. \eqref{ke}, inset). 
\begin{figure}[htb!]
\includegraphics[width=1\columnwidth]{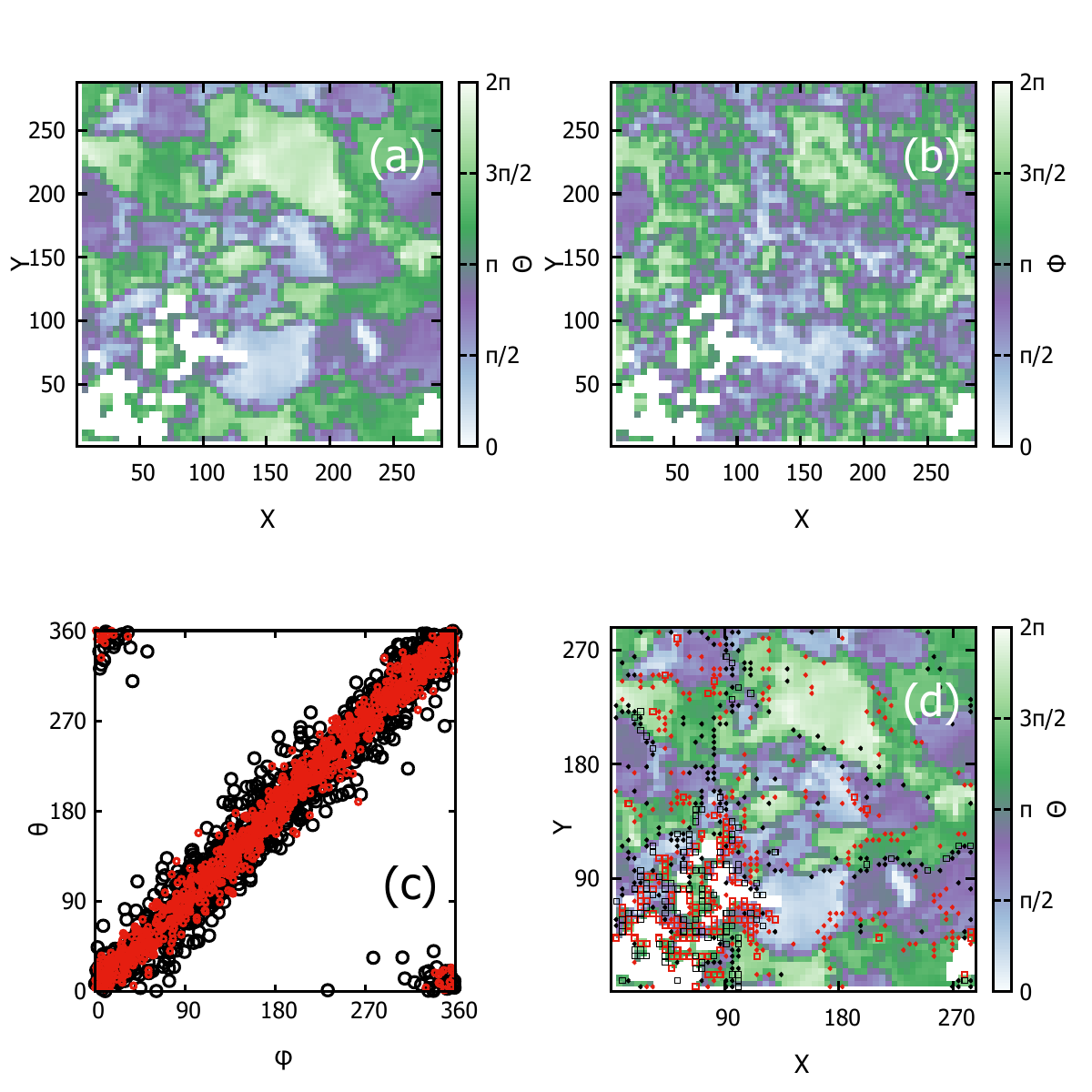}
\caption{Spatial maps of the orientation of (a) coarse-grained velocity {\bf v} given by $\Theta=\tan^{-1} (v_{y}/v_{x})$ and (b) propulsion force {\bf f} given by $\Phi = \tan^{-1} (f_{y}/f_{x})$ coarse-grained over a scale $\Omega=4.73$ for $N=10^5$. Note that these maps have been interpolated for the sake of visual presentation. (c) Correlation between the orientation $\theta$ of the time averaged velocity of a particle and the orientation $\phi$ of the propulsion force acting on it.
(d) Spatial map of coarse grained $\Theta$, overlapped with regions with with positive (in red) and negative (in black) vorticity, $\omega = \nabla \times {\bf v}$. High ($|\omega| > 0.15$) and intermediate ($0.15 > |\omega| > 0.08$) values, are shown in squares and circles, respectively, and are seen to be enriched at the interfaces of the velocity domain, \textcolor{black}{suggesting that topological changes are driven by the interplay between dynamics of emergent cavities and evolution of the orientational structures.} }
\label{orientation}
\end{figure}


\begin{figure}[htb!]
\includegraphics[width=0.9\columnwidth]{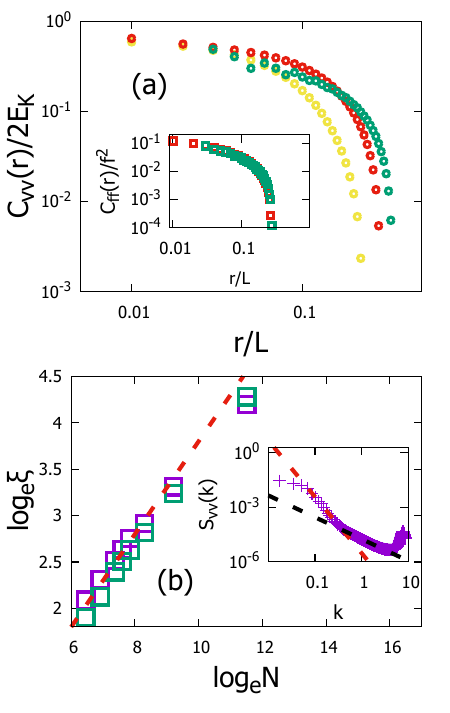}
\caption{(a) Normalized velocity and force correlation functions $C_{vv}(r)$ vs. scaled distance $r/L$
for $N=10^3$ (green), $N=10^4$ (green) and $N= 10^5$ (yellow). Inset. $C_{ff}(r)/f^2$ vs $r/L$ for $N=10^3$ (green), $N=10^4$ (green)
(b) Correlation lengths, $\xi$ extracted from $C_{vv}(r)$ (green) and $C_{ff}(r)$ (purple), plotted versus $N$ in logarithmic scale. The dashed line has a slope 0.5. 
Inset. The structure factor $S_{vv}(k)$ for $N=10^5$ plotted versus $k$ in logarithmic scale, with fits to power laws for two ranges of values of $k$.}

\label{corr_func}
\end{figure}
\begin{figure}[htb!]
\includegraphics[width=0.8\columnwidth]{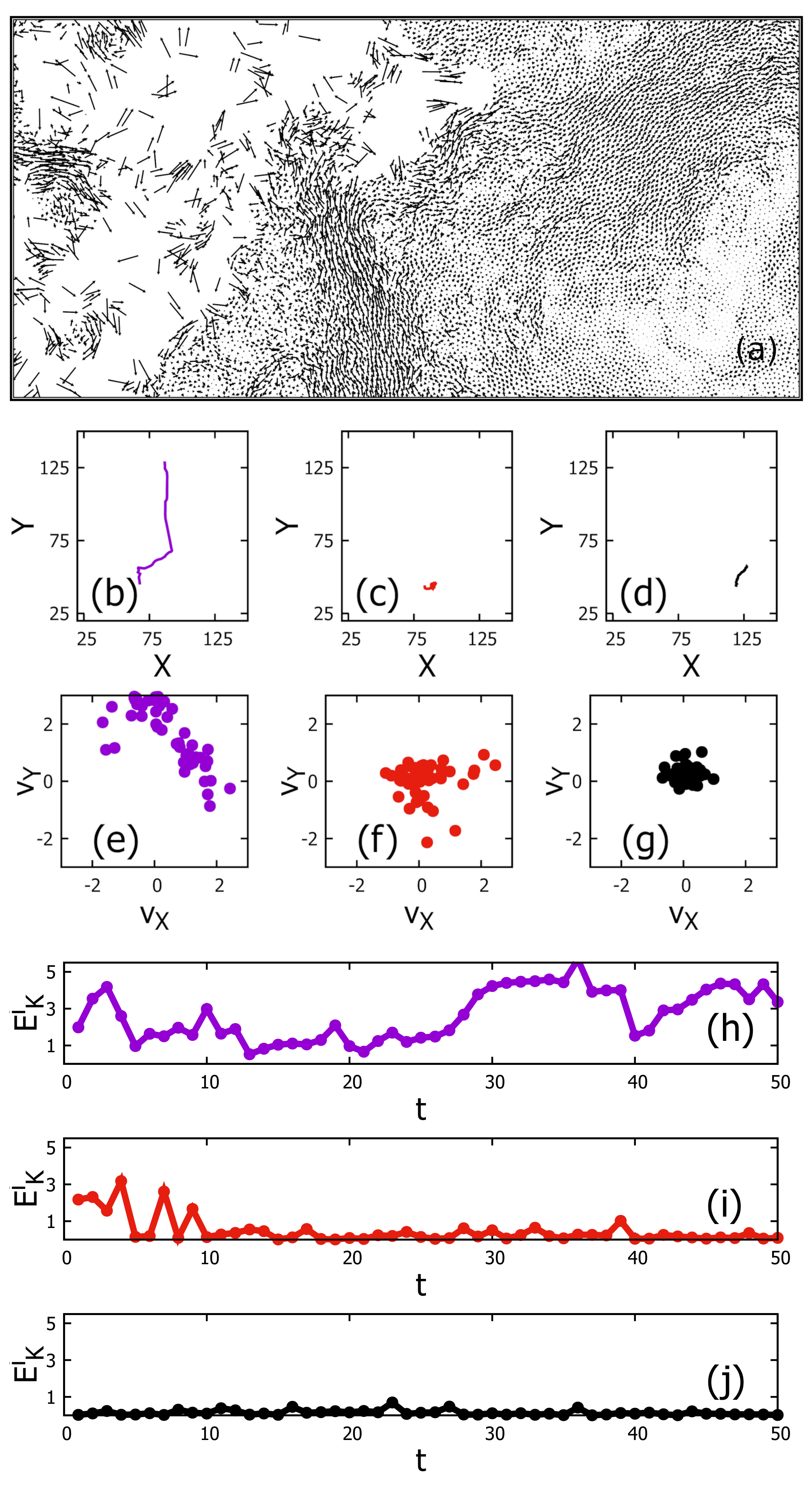} 
\caption{Dynamical heterogeneity: (a) Snapshot of a region showing simultaneous presence of mobile co-moving clusters and slowly varying regions around the cavity within a time interval of $t=50\tau_{LJ}$. 
Magnitude of instantaneous displacements are shown by arrows. Single-particle Tracking shows three distinct types of motion --- (A) Fast (B) Intermittent (C) Slow. 
Representative (b-d) particle trajectories, 
(e-g) Correlation of velocity components and (h-j) kinetic energy time-series are shown at a single particle level. The corresponding movie could be seen in SI Movie.1. Associated time-series of global kinetic energy at a much larger time interval is shown in SI Fig. 6. See SI Note 6 and SI Figs. 8-9 for related discussion.}
\label{dyn_hetero}
\end{figure}

We have also calculated the velocity structure factor, \textcolor{black}{$ S_{vv}({\bf{k}}) = \langle{\bf{v}}({{\bf{k}}})\cdot{\bf{v}}(-{\bf{k}})\rangle$
where $\langle\cdots\rangle$ represents an average over different times in the steady state and different runs, and $ {\bf{v}}({\bf{k}}) = \frac{1}{N}\sum_{j=1}^{N} \exp[i {\bf{k}} \cdot {\bf{r}_j}] {\bf{v}}_j$.} We circularly average $S_{vv}({\bf{k}})$ over $\bf{k}$ to obtain  $S_{vv}(k)$ where $k = |{\bf{k}}|$. Fig. 2 Inset. shows a plot of $S_{vv}(k)$ versus $k$ in logarithmic scale. There are two ranges of $k$-values in which the $k$-dependence of $S_{vv}(k)$ can be represented by power laws, $S_{vv}(k) \propto k^\alpha$. We find that $\alpha \simeq 3$ in the smaller $k$ region. This is in agreement with Porod's law~\cite{porod83} of domain growth, $S(k) \propto k^{d+1}$ where $d$ is the spatial dimension, which signifies the occurrence of sharp domain boundaries. The origin of the exponent found in the larger-$k$ range, $\alpha \simeq 1.2$, is not clear. It may be related to the turbulent behavior discussed below.

 The sorting into velocity domains is accompanied by significant {\it dynamical heterogeneity} (Fig.4(a)). Particles in the low density regions move ballistically over fairly long time windows and those in the high density bulk move with substantially lower speeds, while some are kinetically arrested (Fig. 4(b-j) (also see SI Figs. \eqref{dh_time} and \eqref{dh_map}. 

At late times, the movie of the particle dynamics in the $N=10^5$ system, suggests a kind of turbulence, whose origin  is different from the usual low-Reynolds number turbulence reported in active nematics \cite{alert, giomi, yeomans}. Here, turbulence in this low-Reynolds number compressible active fluid is driven by the motion of the large dynamic voids or cavitation bubbles. 
The turbulence is reflected in the time series of the displacement or the kinetic energy of a typical particle which shows intermittent behavior, \textcolor{black}{facilitated by cascade of plastic rearrangements, due to topological stress build-up and its redistribution at the neighboring regions}~\cite{footnote1, suman2}.

In conclusion, we have shown that a persistently active fluid at high forcing displays unusual properties compared to its passive counterpart.
The development of velocity and force sorting across all system sizes studied is the most striking aspect of our study.
The velocity sorting is likely to be due to a tendency to reduce the power dissipation $\frac{\eta}{2} \int |\nabla {\bf v}|^2 \, d^{2}x$, where $\eta$ is the kinematic viscosity of the fluid~\cite{geigenfeind2020}. This will be investigated in
 a later study, where we will analyze the time development of these co-moving and co-aligned domains using a 
coarsening picture~\cite{chandan2}. The growth of the correlation length of these domains with system size is possibly cutoff by the formation and motion of macroscopic voids that drive a novel turbulence, \textcolor{black}{ with an elasto-plastic origin}~\cite{suman3}. 

\section{Acknowledgements}

S. D. acknowledges the support by the Department of Atomic Energy, Govt. of India Projects No. RTI4001 and RTI4006. All the simulations were carried out in the {\it Tetris} cluster of the HPC Unit at the ICTS-TIFR. S. D. also acknowledges the Indian Institute of Science (IISc) for a visiting fellowship and support of Institute of Mathematical Sciences (IMSc). We acknowledge insightful discussion with L. Berthier, S. Ramaswamy and P. Sollich. M. R. acknowledges support from the Department of Atomic Energy (India) under project no.\,RTI4006, the Simons Foundation (Grant No.\,287975), and DST (India) for a J.C. Bose Fellowship. C.D. acknowledges support from SERB (India) for a SERB Distinguished Fellowship.

{\it Research Contributions.---} C. D.: Conceptualization (Part), Discussion, Comments, Validation, Presentation, Review and Editing; M. R.: Conceptualization (Part), Discussion, Data Interpretation (Part), Review and Editing; P. C.: Discussion and Review; S.D.: Conceptualization (Part), Framework development, Simulations and Analysis of Data, Visualizations, Data Interpretation (Part), Presentation, Preparation of the First draft, Review and Editing, Project Coordination. 

\vspace{15in}

\appendix

\section{1. Details of Model and Simulations}

\setcounter{figure}{0}

\begin{figure}[h]
\includegraphics[width=0.7\columnwidth]{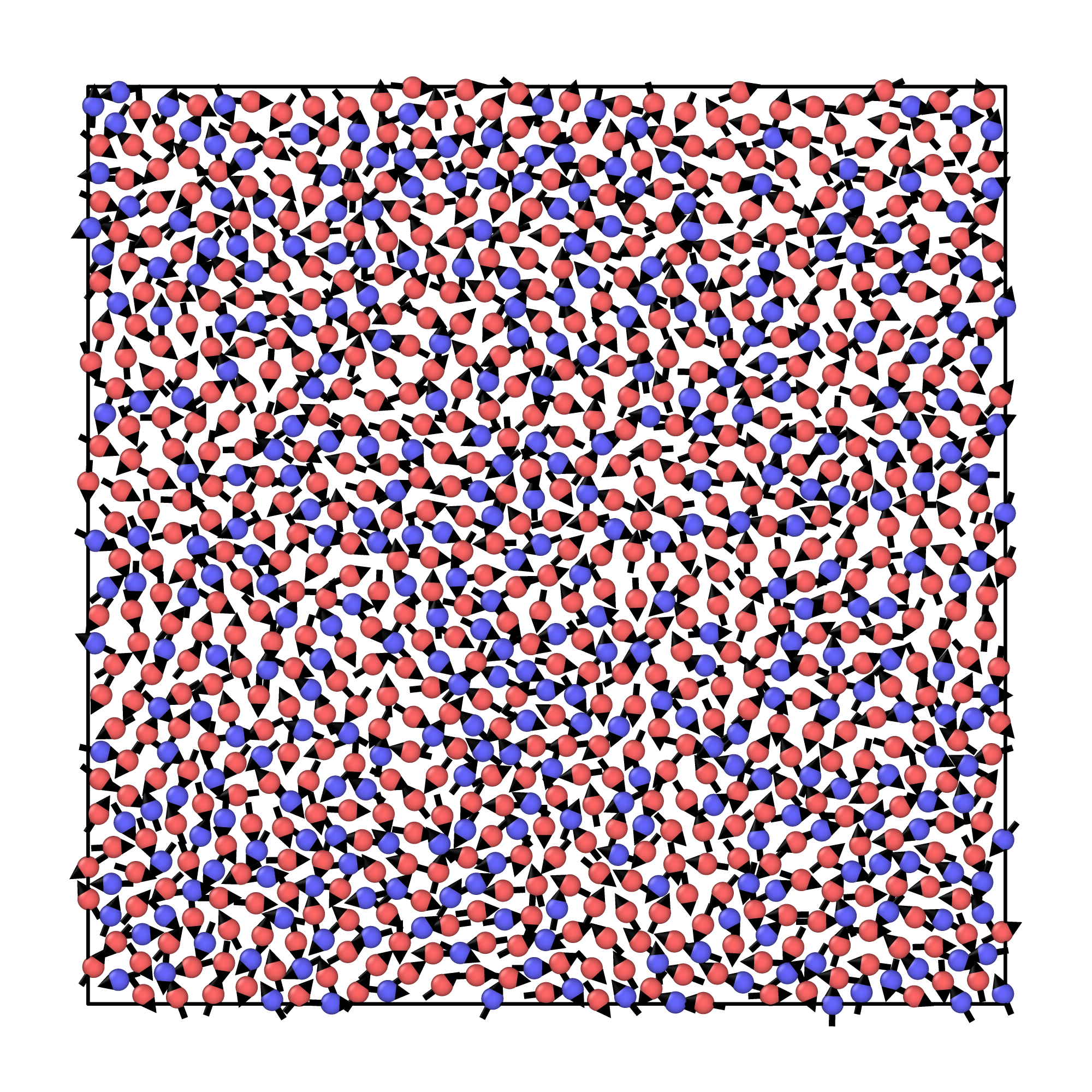}
\caption{Particle positions are shown for a configuration in the steady state for $N=10^3$. Different species are shown in different colors and directions of the self propulsion forces are indicated by the arrows.}
\label{config}
\end{figure}

This work is based on Langevin simulations of an athermal binary mixture of Lennard-Jones particles \cite{koboriginal} of mass $m$, subjected to random persistent self-propulsion forces at particle scale in two spatial dimensions, \cite{rituporno2,dam8}: 
\begin{equation}
   m\ddot{\bf{r}_{i}}=-\Gamma\dot{\bf{r_{i}}}+\sum^{N}_{j \ne i=1}\bf{f}_{ij}+\bf{f}_{i} \nonumber 
\end{equation}
with ${\bf{f}_{i}}=f\hat{n}_{i}$ where $\hat{n}_i \equiv (\sin{\theta}_i, \cos{\theta}_i)$ assigns the direction of the random persistent self-propulsion force for a particle $i$ at position ${\bf{r}}_{i}$ with $\sum_{i=1}^{N}{\bf{f}}_{i}= 0$, in a medium with damping $\Gamma$. Simulations are performed using LAMMPS \cite{lammps}. The binary mixture consists of two different types of particles $A$, $B$ with a proportionality of $65:35$. Any pair of such particles of mass $m(=1)$, $i$ and $j$ of types $\alpha(\in [A,B])$, $\beta(\in [A,B])$ at a separation $r$ $( = |{\bf{r}_{i}-\bf{r}_{j}}|<r^{\alpha \beta}_{C})$ interact at a constant density $\rho(=1.2)$ via the Lennard-Jones potential given by  
\begin{equation}
V^{(i,j)}_{\alpha \beta}(r)=4\epsilon_{\alpha \beta} \left[\left(\frac{\sigma_{\alpha \beta}}{r}\right)^{12}-\left(\frac{\sigma_{\alpha \beta}}{r}\right)^{6}\right]
\nonumber
\end{equation}
We use $\sigma_{AB}=\sigma_{BA}=0.8\sigma_{AA}$, $\sigma_{BB}=0.88\sigma_{AA}$, $\epsilon_{AB}=\epsilon_{BA}=1.5\epsilon_{AA}$, $\epsilon_{BB}=0.5\epsilon_{AA}$, truncated at a threshold $r={r}^{\alpha \beta }_{C}(=2.5{\sigma}_{\alpha \beta})$ ensuring that the potential and its first derivative continuously goes to zero at the cutoff \cite{lammps}. $\sigma_{AA}$ is used as the unit of length and the unit of time is set at $\tau_{LJ}=\sqrt{m\sigma^{2}_{AA}/\epsilon_{AA}}$. We use system sizes between $N=L^2 =10^2-10^5$. A typical configuration in the steady state for $N=10^3$ is shown in Fig.\ref{config} of the SI.

  \section{2. Static Structure}
\begin{figure}[h]
\includegraphics[width= 0.7\columnwidth]{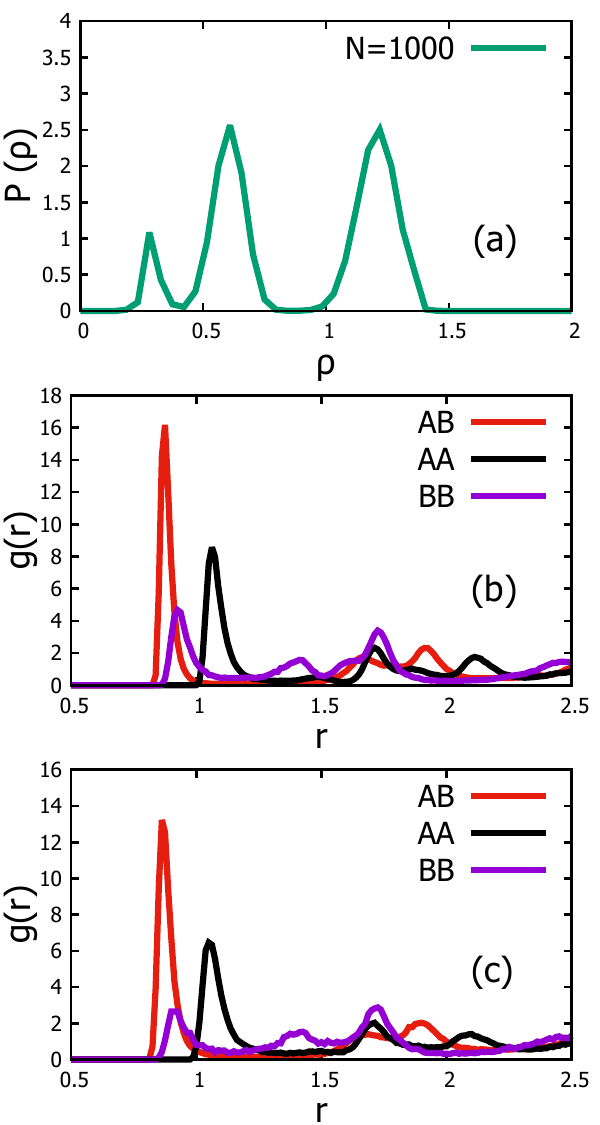}
\caption{Steady-state Structures. (a) Distribution of the local density $\rho$ in the steady state of a system with $N=10^3$.  Partial Radial Distribution Functions, $g^{\alpha \beta}(r)$ for $N=10^3$ for $\alpha,\beta \in [A,B]$ for (a) a passive state ($f=T=0$) and (b) active state ($f=3$). $g^{\alpha \beta}(r)$ for different combinations of $\alpha,\beta$ indicate amorphous structures in both active and passive limits.} 
\label{rdf}
\end{figure}

  Fig.\ref{rdf}, panel (a) shows the distribution of the local density $\rho$ in the steady state of a system with $N=10^3$. The distribution shows 3 peaks, as in panel (b) of Fig. 1 of the main text which shows the distribution of $\rho$ for a larger system with $N=10^5$. However, the positions of the peaks for the two values of $N$ are different. In particular, the first peak appears at a non-zero value of $\rho$ for the smaller system, indicating the absence of voids.
  
  We investigate the structures of such liquid states by computing the partial radial distribution functions in real-space. For any two particles of species $\alpha$ and $\beta$, separated at a distance $r_{ij}^{\alpha\beta}$, such partial radial distribution functions are given by
  \begin{equation}
      g^{\alpha \beta}(r)=\frac{A}{N^{\alpha}N^{\beta}}< \sum_{i=1}^{N^{\alpha}} \sum_{j=1}^{N^{\beta}} \delta (r-r^{\alpha \beta}_{ij})>. \nonumber
  \end{equation}
  where $A$ is chosen to ensure that $g^{\alpha \beta}(r) \to 1$ for large $r$.  In Fig. \ref{rdf}, we show $g^{\alpha \beta}(r)$ for $N=10^3$ and $f=3$. The partial structures are neither significantly different from the structures obtained in the passive athermal limit nor very different for different $N$ (Data not shown). With increasing $N$, we only note some marginal changes, such as small shifts of the second and third peaks towards smaller values of $r$. 
  
\section{3. Number Fluctuations}
\begin{figure}[h]
\includegraphics[width=1\columnwidth]{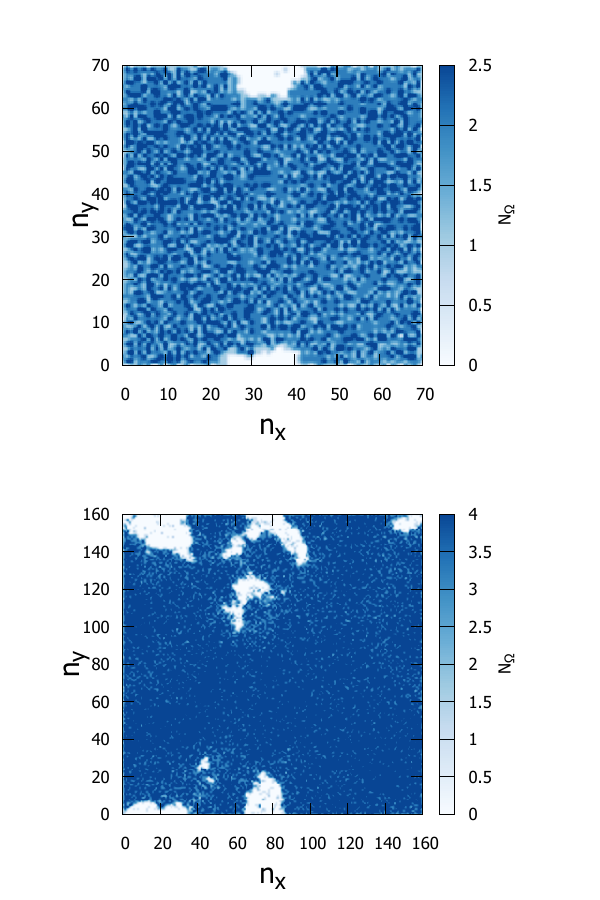}
\caption{Typical spatial maps of the number density for (a) $N=10^4$ (b) $N=10^5$ obtained at scales $\Omega=1.285$ and $\Omega=1.793$ respectively. Prominent cavities could be seen with rough interfaces for both the system sizes.}
\label{number_density}
\end{figure}

\begin{figure}[h]
\includegraphics[width=0.9\columnwidth]{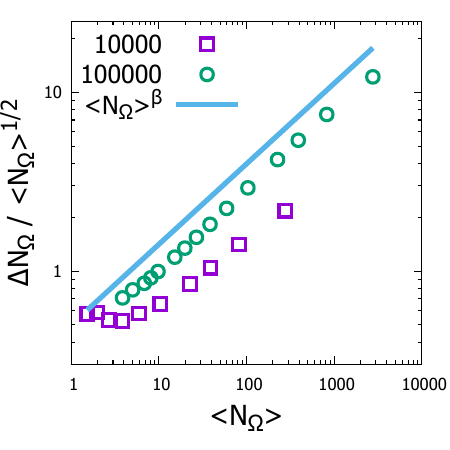}
\caption{Giant number fluctuations for two different system sizes, $N=10^4$ and $N=10^5$. The data was obtained using statistics from $10-50$ independent trajectories. The solid line corresponds to $ <N_{\Omega}>^{\beta}$ with $\beta\approx 0.45$. This line is only a guide to the eye.}
\label{giant_number}
\end{figure}

We divide the system into multiple boxes of size $\Omega$. In every such box, we track the evolving number of particles, $N_{\Omega}(t)$ in the steady state. We calculate the steady-state mean, $<N_{\Omega}>$, and the standard deviation with respect to it, $\Delta N_{\Omega}=\sqrt{<N_{\Omega}^2>-<N_{\Omega}>^2}$ for a given $\Omega$, using the time series, $N_{\Omega}(t)$. The number fluctuation is given by $\Delta N_{\Omega}/{\sqrt{<N_{\Omega}>}}$. We average it over different spatial regions and also over different independent trajectories. We show typical maps of the local number density in SI Fig.\ref{number_density}. Respective data for the number fluctuation is shown in SI Fig.\ref{giant_number}. They show that the slope of the $\Delta N_{\Omega}/{\sqrt{<N_{\Omega}}>}$ vs. $<N_{\Omega}>$ plot at the large-$N_\Omega$ limit is similar for the two values of $N$, whereas the dependence for smaller values of $<N_\Omega>$ seems to have some differences between $N=10^4$ and $N=10^5$.  

  \section{4. Correlation Functions}

  We compute the correlation functions between the velocities and self propulsion forces at single particle level for any two particles $i$ and $j$ located at ${\bf r}_{i}$ and ${\bf r}_{j}$ respectively, defined as: 

  \begin{equation}
      C_{\mu \nu}(r)= \frac{\sum_{i=1}^{N-1}\sum_{j=i+1}^{N} ({\bf \mu}_{i}\cdot{\bf \nu}_{j}) \delta(|{\bf r}_{i}-{\bf r}_{j}|-r)}{\sum_{i=1}^{N-1}\sum_{j=i+1}^{N} \delta(|{\bf r}_{i}-{\bf r}_{j}|-r)} \nonumber
  \end{equation} 
with ${\bf \mu}_{k},{\bf \nu}_{k} \in ({\bf v}_{k},{\bf f}_{k})$ where $k \in (1,N)$. We also average $C_{\mu \nu}(r)$ over independent trajectories. 

\begin{figure}[h]
\includegraphics[width=0.9\columnwidth]{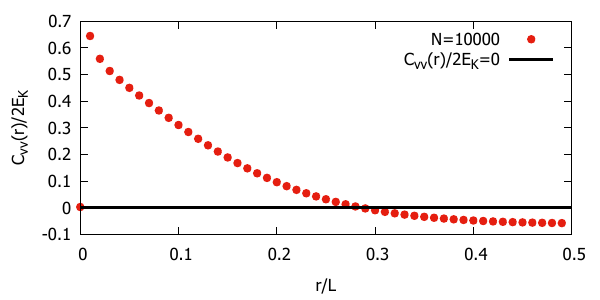}
\caption{Normalized velocity correlation function $C_{vv}(r)/2E_{K}$ vs $r/L$ for $N=10^4$. The solid line indicates $C_{vv}(r)=0$. }
\label{cor}
\end{figure}

We calculate the correlation length ($\xi$), by identifying the first zero of $C_{\mu \nu}(r)$, that satisfies $C_{\mu \nu}(r=\xi)=0$. This is illustrated in SI Fig. \ref{cor}.

\section{5. Energetics of the Steady State}

We compute the kinetic energy at single particle level from the single particle velocities, ${\bf v}_i$, obtained by integrating the equation of motions with time step $\delta t=0.005\tau_{LJ}$. We compute the mean kinetic energy per particle from a single configuration by computing, $E_{K}=\frac{1}{2N}\sum_{i=1}^{N}|{\bf v}_i|^{2}$. We then calculate $<E_{K}>$ where $<\ldots>$ implies an average over both configurations in the steady states in a trajectory, and over multiple trajectories obtained from different origins. 

Typical time-series for $E_K$ are shown in SI Fig. 6. The dependence of the mean kinetic energy $<E_K>$ on $N$ is shown in SI Fig. 7.
From the data of $E_{K}$, we compute the probability distribution of the system to have a kinetic energy $E_{K}$, $P(E_{K})$, and we identify the mode, $\mu^{P}=\max_{E_{K}} P(E_{K})$. The peak of the distribution of $P(E_{K})$ is shifted to zero, to obtain the rescaled distribution, $P(E_{K}-\mu_{P})$. We observe that $P(E_{K}-\mu_{P})$ is non-Gaussian and has extended tail with positive skewness (Inset of SI Fig. 7). 

\begin{figure}[h]
\includegraphics[width=\columnwidth]{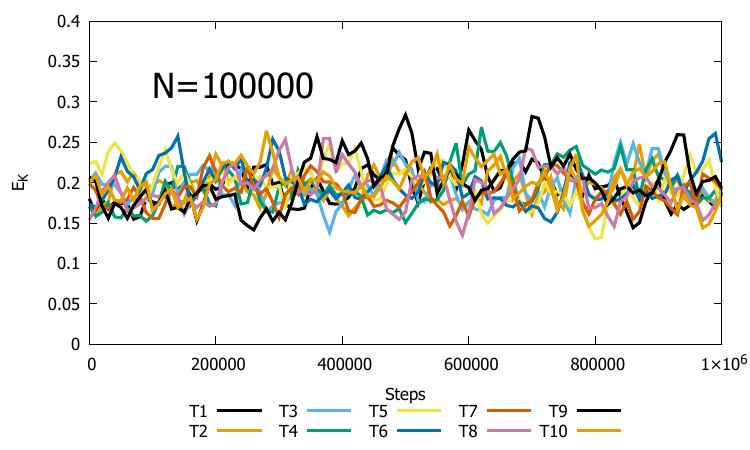}
\caption{Kinetic Energy time series: Time evolution of the kinetic Energy per particle, $E_{K}$ for $N=10^5$, from ten different trajectories obtained from independent origins.} 
\label{ss_ke}
\end{figure}

\begin{figure}[h]
\includegraphics[width=0.7\columnwidth]{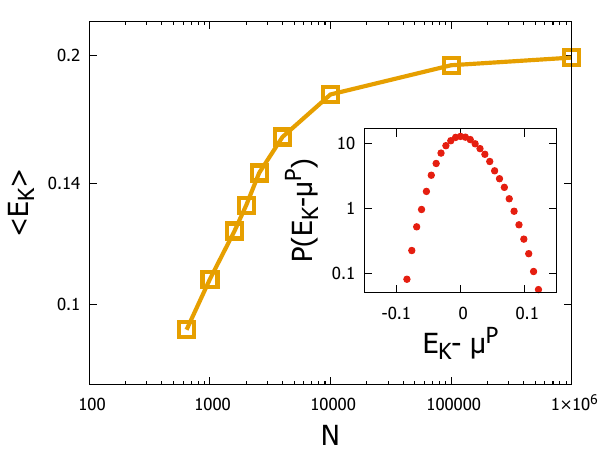}
\caption{Steady state energetics: Dependence of the mean kinetic Energy per particle, $<E_{K}>$ (Yellow squares) on $N$. Inset: $P(E_{K}-\mu^{P})$ vs $E_{K}-\mu^{P}$ for $N=10^4$. } 
\label{ke}
\end{figure}

\section{6. Dynamic Heterogeneity}
\begin{figure*}[h]
\includegraphics[width=2\columnwidth]{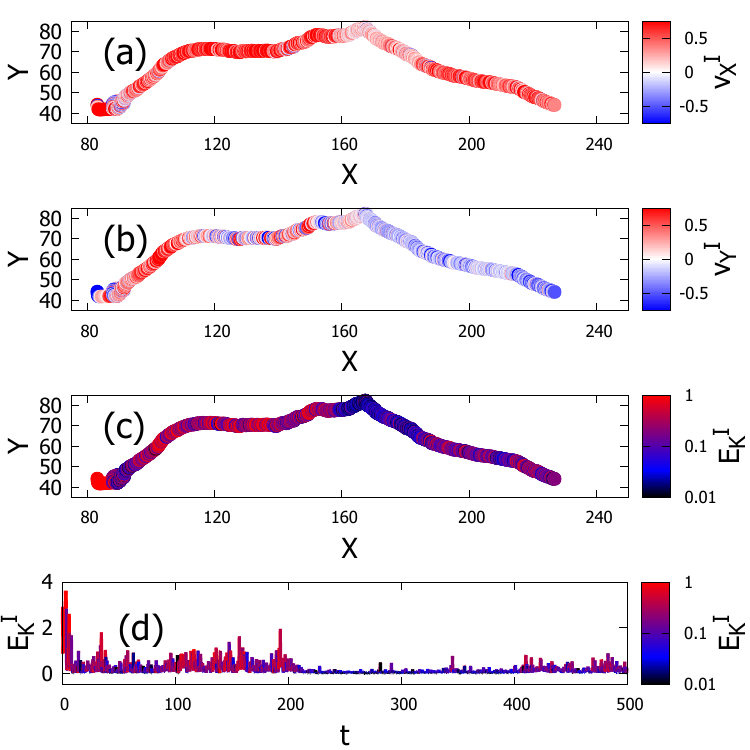}
\caption{Single Particle Tracking $(N=10^5)$. Dynamics of a typical particle undergoing dynamic heterogeneity: Spatial map of the particle trajectory in which the particle has been color coded by its values of (a) $v_{X}^{I}$ and (b) $v_{Y}^{I}$ and (c) Kinetic energy $E_K^I$. Associated time-series of the data for $E_{K}^{I}$ is shown in (d) for an interval of $500\tau$ shows that the dynamics can become locally fast, intermittent or slow, depending on its spatiotemporal environment. } 
\label{dh_time}
\end{figure*}

In order to probe the presence of heterogeneity within the particle dynamics, we show the dynamical behavior of a single particle over a sufficiently long interval that shows distinct behavior in the course of its motion. This we illustrate in SI Fig. \ref{dh_time} (a-d). The particle shows large $E_{K}$ in the initial period while it slows down with time and remains caged for a long interval after which it becomes fast again.

The particle dynamics in Fig. \ref{dh_time} falls in one of the 3 classes --- I. fast II. intermittent and III. Slow, depending on the spatiotemporal environment it explores in the course of its dynamics. 

\begin{figure*}[h]
\includegraphics[width=1.05\columnwidth]{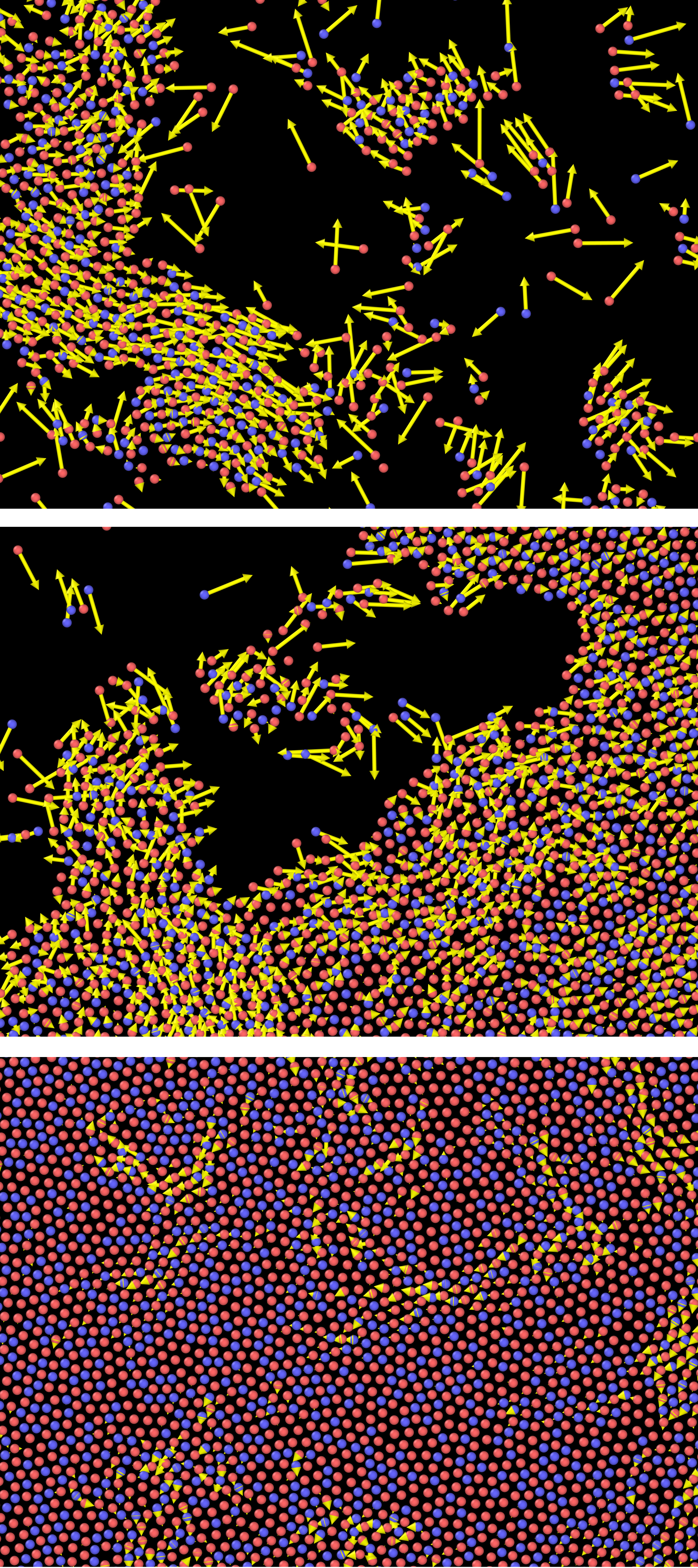}
\caption{Dynamic Heterogeneity: Single particle displacement maps obtained at a particle scale shows presence of heterogeneity in dynamics ($N=10^5)$. The displacements are shown by the arrows and the magnitudes of the displacements are proportionately given by the lengths of the arrows. The three types of dynamics -- (A) fast, (B) intermittent, and (C) slow are shown in the three snapshots (top, middle and bottom, respectively). The regions are dominated by (A) co-moving clusters (B) interfacial dynamics and (C) dynamically localized particles for which displacements are mediated by plastic rearrangements that are clearly seen in the bottom plot.}
\label{dh_map}
\end{figure*}

We further illustrate the spatial dynamics in Figs. \ref{dh_map} in which we show that these three distinct classes appear in the dynamics because of the presence of I. co-moving clusters, II. interfacial dynamics and III. localized motion. 

\section{7. Velocity Relaxation}
\begin{figure*}[h]
\includegraphics[width=1.2\columnwidth]{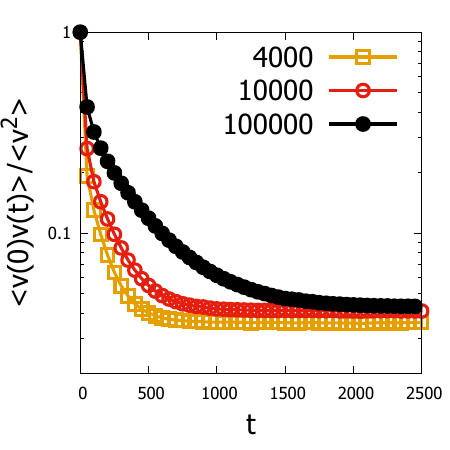}
\caption{Velocity Relaxation: Normalized velocity auto-correlation function, $\frac{<{\bf v}(0)\cdot {\bf v}(t)>}{<|{\bf v}|^{2}>}$ for different system sizes, ($N=4\times10^3,10^4,10^5$. They saturate at non-zero values at long $t$, suggesting a non-zero time average.} 
\label{vacf}
\end{figure*}

We compute the velocity autocorrelation function (VACF), $\langle {\bf v}_i(0) \cdot {\bf v}_i(t)\rangle$ where ${\bf v}_i$ is the velocity of the $i^{th}$ particle, and average it over the time origin and all the particles. As shown in  Fig.\,\eqref{vacf}) for $N=10^5$, the VACF saturates to a finite value at large  $t$. This is an indication of a finite time-average of the velocity which, as shown in Fig. 2(c) of the main text, points in the direction of the self-propulsion force acting on the particle. The time averaged velocity increases with $N$ and appears to saturate for large $N$.
This can be understood from considering the motion of a single particle under the action of the constant self-propulsion force acting on it. Interactions of this particle with the other particles in the system provide a bath with friction and a random noise. The motion of this particle can then be modeled by that of the Brownian particle in the presence of a constant external force, which leads to a constant average velocity in the direction of the external force.


\begin{thebibliography}{99}
\bibitem{footnote1} \textcolor{black}{We make sure that this is not an effect of the periodic boundary conditions, by noting that the time scales over which intermittency is observed is smaller than the transit  times of a typical particle to traverse the box. We leave a detailed characterization of this unusual turbulence for a later study.}
\end{thebibliography}
\end{document}